\documentclass[preprint,aps,prd,superscriptaddress]{revtex4-1}
\usepackage{hyperref}
\usepackage{graphics}
\usepackage{color}
\usepackage{amssymb}
\usepackage{graphics, slashed}
\usepackage{amsmath}
\usepackage{slashed}
\usepackage{chngcntr}
\usepackage[normalem]{ulem} 
\usepackage {ulem} 
\usepackage{colordvi}
\def\qmax{q_{max}}
\def\ptmin{p_{tmin}}
\begin{document}
\title{
\bf   Inelastic cross-section and
survival probabilities at LHC in mini-jet models}
\date{\today}
\author{ Daniel A. Fagundes}
\email{daniel.fagundes@ufsc.br}
\affiliation{Departamento de Ci\^{e}ncias Exatas e Educa\c{c}\~{a}o, Universidade Federal de Santa Catarina -  Campus Blumenau, 89065-300, Blumenau, SC, Brazil. }
\author{Agnes  Grau}
\email{igrau@ugr.es}
\affiliation{Departamento de Fisica Teorica y del Cosmos,
Universidad de Granada, 18071 Granada, Spain}
\author{ Giulia  Pancheri}
\email{pancheri@lnf.infn.it}
\affiliation{INFN Frascati National Laboratories, 00044 Frascati, Italy}
\altaffiliation{Also affiliated to Center for Theoretical Physics, MIT, Cambridge,  MA 02139, USA}
\author{Olga  Shekhovtsova}
\email{Olga.Shekhovtsova@lnf.infn.it}
\affiliation{NSC KIPT, Kharkov, 61108, Ukraine, 
and 
INP of PAS, Cracow, 31-234 Poland}
\author{Yogendra N. Srivastava}
\email{yogendra.srivastava@gmail.com} 
\affiliation{Department of Physics \& Geology, University of Perugia, 06123 Perugia, Italy}
\affiliation{Physics Department, Northeastern University, Boston, Mass 02115, USA}
\begin{abstract}

Recent results for the total
 and inelastic hadronic cross-sections from LHC experiments 
are compared with predictions from a single channel   PDF driven eikonal mini-jet model and from an empirical model.
 The role of soft gluon resummation in the 
infrared region in taming the rise of mini-jets and  their contribution to the  increase of the total cross-sections at high
energies are discussed. 
Survival probabilities at LHC, whose theoretical estimates 
range  from  circa 10\% to a few {\it per mille},  will be estimated in this model and compared with 
results from QCD inspired models and from multichannel eikonal models. 
We revisit a previous calculation and examine  the origin of 
these discrepancies.

 \end{abstract}
\preprint{INFN-17-12/LNF, MIT-CTP/4910 }
\maketitle
 \section{Introduction}
In this paper we  present an estimate  of survival probabilties in hadronic collisions,  
obtained with  the eikonal mini-jet model first proposed in   \cite{Durand:1988ax}, and later  
implemented with soft gluon resummation  in \cite{Corsetti:1996wg,Grau:1999em,Godbole:2004kx}. 
 We shall make use of latest measurements
 by the TOTEM Collaboration  \cite{Antchev:2013iaa}, 
 at 7 TeV for all 3 cross-sections, and  at  8 TeV 
  in the Coulomb region and with luminosity independent measurements   \cite{Antchev:2016vpy,Antchev:2013paa}, by  
 CMS \cite{Chatrchyan:2012nj}  and LHCb  \cite{Aaij:2014vfa}  for the inelastic cross-section at 7 TeV, by the ALICE Collaboration for the inelastic cross section at 2.76 and 7 TeV \cite{Abelev:2012sea},
  by the   ATLAS  Collaboration 
 for the  total, inelastic and elastic $pp$ cross-sections at 7 \cite{Aad:2014dca} and 8 TeV \cite{,Aaboud:2016ijx},   and by measurements of the inelastic part at $13$ TeV by CMS 
\cite{VanHaevermaet:2016gnh} and ATLAS \cite{Aaboud:2016mmw}.
 
Survival probabilities were originally discussed in \cite{Dokshitzer:1991he,Bjorken:1992er} 
to estimate the probability associated 
with a hard process when no low transverse momentum 
particle production is present 
{in the central region}. In  \cite{Bjorken:1992er}, such a probability  
was estimated to be {around}  $5 \%$ at the Superconducting Super Collider SSC ($\sqrt{s}=40$ TeV), 
but with an overall possible uncertainty of a factor three in either direction. 
Presently, for LHC 
{data up to  $\sqrt{s}=13$ TeV}, estimates  vary between those of  a QCD inspired model
\cite{Block:2015sea} where   the survival probability is calculated to be   13 \%, to  calculations within 
the 
{Regge-Pomeron approach} 
which range between $(0.7\div 2) \%$ in \cite{Khoze:2013dha} and between 
$(0.25\div 3) \%$  in \cite{Gotsman:2015aga}. 
Such large discrepancies  
arise due to (i) the choice of the 
impact parameter distribution of partons involved in the scattering and, to a lesser extent, to (ii) the 
estimate of the inelastic total cross-section. 
{As for  data on rapidity gaps, LHC measurements at 7 TeV by ATLAS  \cite{Aad:2015xis} and CMS \cite{Chatrchyan:2012vc} Collaborations are affected by rather large errors and cannot yet discriminate between models.}
   
In the following, in the quest for a clearer definition of survival probabilities (SPs), we shall 
employ eikonal mini-jet models 
to clarify and sharpen the physical meaning of the survival probability concept.
Comparison with other models will also be 
made.

{Mini-jets were first introduced in estimates of hadronic physics in \cite{Horgan:1980fm,Pancheri:1985sr,Sjostrand:1987su,Durand:1988ax} but were not yet recognized  as dominant in proton-proton collisions when the earlier estimates 
of 
SPs
appeared \cite{Bjorken:1992er}.  
Since then a  better understanding of the role played by mini-jets in high energy collisions has been
achieved, 
 including proposal for  beyond the leading power calculations \cite{Kotko:2016lej}.}

In the following, after    a brief summary of the main  features  of the  PDF (Parton Density Function) driven
mini-jet model  
that we employ,   we  examine   the  most recent data for  the total cross-sections, 
and address  the question of the inelastic cross-section in single  channel eikonal models.
We then apply our model to  discuss SPs
for hard and soft distributions of partons in the 
protons and clarify the difference arising from using different impact parameter distributions.

{Revisiting a previous calculation in \cite{Achilli:2007pn}, 
we put forward a new proposal, which reduces the   estimate of $\simeq 10\% $ at LHC energies  by  
almost an order of magnitude. This proposal is based on the physical meaning of the survival probability concept 
in mini-jet models and on explicit inclusion in  the calculation   of the  soft gluon effects accompanying mini-jet processes. 
Our resummation procedure is based on Poisson distributed soft gluon emissions and on an hypothesis of 
maximal singularity of the soft gluon spectrum.}

\section{Accelerator data and the total $pp$ cross-section in a  PDF driven eikonal model}
 We consider an eikonal  model, such as
   \begin{equation}
     \sigma_{total}= 2  \int d^2{\bf b}[1-e^{-\chi_I (b,s)}  ] \label{eq:sigtot}
   \end{equation}
   where  the eikonal function is taken to be purely imaginary 
   at high energies, 
   and contains contributions from  both soft and semi-hard collisions.  
   
   For the imaginary part of the eikonal, $\chi_I(b,s)$,
   we write 
   \begin{equation}
   2 \chi_I(b,s)={\bar n}_{soft}(b,s)+ {\bar n}_{hard}(b,s)
  = A_{FF}(b) \sigma_{soft}(s)+A_{BN}(b,s)\sigma_{mini-jet}(s,p_{tmin})\label{eq:nbs}
   \end{equation}
   where  $2\chi_I(b,s)$ can be seen to correspond to  the average number of Poisson distributed 
   parton-parton collisions \cite{Durand:1988ax,Achilli:2011sw,Fagundes:2015vba}.The distinction between 
   the two terms at the right hand side of  Eq.~(\ref{eq:nbs}) is done on the basis of using a perturbative 
   QCD (pQCD) calculation for the mini-jet cross-section, i.e. for all interacting partons with $p_t    \ge p_{tmin}$ 
   \cite{Grau:1999em}. Namely, $p_{tmin}$ is the  scale 
     of {\cal O}(1-2\ GeV) which  phenomenologically separates collisions between partons 
     {exiting the scattering with final momenta  $p_t>p_{tmin}$, aka mini-jets.}
     
 { Hadronic activity not associated to  mini-jet production can    be included in ${\bar n}_{soft}(b,s)$, such as collisions  
 leading to  final partons with $p_t<p_{tmin}$. However,  notice that the hadronic activity with   partons with 
 $p_t<p_{tmin}$ can come both through ${\bar n}_{hard}(b,s)$ and ${\bar n}_{soft}(b,s)$, because of the 
 soft gluon emission accompanying the hard (mini-jet) processes, as we shall describe below. We should 
 also point out that the two-component separation of  Eq.~(\ref{eq:nbs}) misses to include Single Diffraction, 
 which has an energy dependence different from the mini-jet cross-section. We shall return to this point later in the paper.}   
     
             The  term, 
   $ {\bar n}_{hard}(b,s)$, is obtained from QCD, with the distribution $A_{BN}(b,s)$ to describe  the 
   contribution of soft gluon emission accompanying collisions between partons with final momenta  $p_t>p_{tmin}$.
   The subscript $BN$ refers to our choice of exploiting the full range of soft gluon momenta, down to $k_t=0$, 
   in the spirit of the Bloch and Nordsieck description of soft quanta emission in QED \cite{Bloch:1937pw}. 
   Our application  to QCD has been described in a number of previous publications, starting from  \cite{Nakamura:1983xp} 
   until recently in \cite{Fagundes:2015vba},   
   where we provide details about our calculation of ${\bar n}_{hard}(b,s)$. In Eq.~(\ref{eq:nbs}) both $A_{FF}(b)$ and $A_{BN}(b,s)$ are normalized to 1. 
        
Together with soft gluon resummation, to which we shall turn shortly, the  distinctive element of our  model is that   the mini-jet cross-section is  not 
 parametrized but calculated (at Leading Order (LO)) from the QCD standard expression, and with standard 
 PDFs, DGLAP evoluted,  $f_{i|A}(x_1,p_t^2)$, i.e. 
    \begin{equation}
\sigma^{AB}_{\rm jet} (s;p_{t_{min}}) = \int_{p_{tmin}}^{\sqrt{s}/2} d p_t \int_{4
p_t^2/s}^1 d x_1 \int_{4 p_t^2/(x_1 s)}^1 d x_2 
\sum_{i,j,k,l}
f_{i|A}(x_1,p_t^2) f_{j|B}(x_2, p_t^2)~~
 \frac { d \hat{\sigma}_{ij}^{ kl}(\hat{s})} {d p_t}.
\label{eq:sigjet}
\end{equation}
 with  $i, \ j, \ k, \ l$ to  denote 
the partons and $x_1,\ x_2$ the
fractions of the parent particle momentum carried by the parton.
$\sqrt{\hat{s}} = \sqrt{x_1 x_2 s}$,   $\hat{ \sigma}$ are the
center of mass energy of the two parton system and the hard parton
scattering cross--section respectively. Following the argument given above, 
this expression sums only collisions with outgoing partons of momentum  with $p_t>p_{tmin}$, 
where $p_{tmin}$ is defined as the region of validity of perturbative QCD, i.e. the coupling is 
given by the asymptotic freedom expression  for running  $\alpha_s(p^2_t)$. 
{When the cut-off $p_{tmin}\gtrsim 1-2$ GeV,  it is usual to refer to these type of  processes as {\it mini-jets} \cite{Horgan:1980fm}.}

\begin{figure}
\resizebox{1.0\textwidth}{!}{
\includegraphics{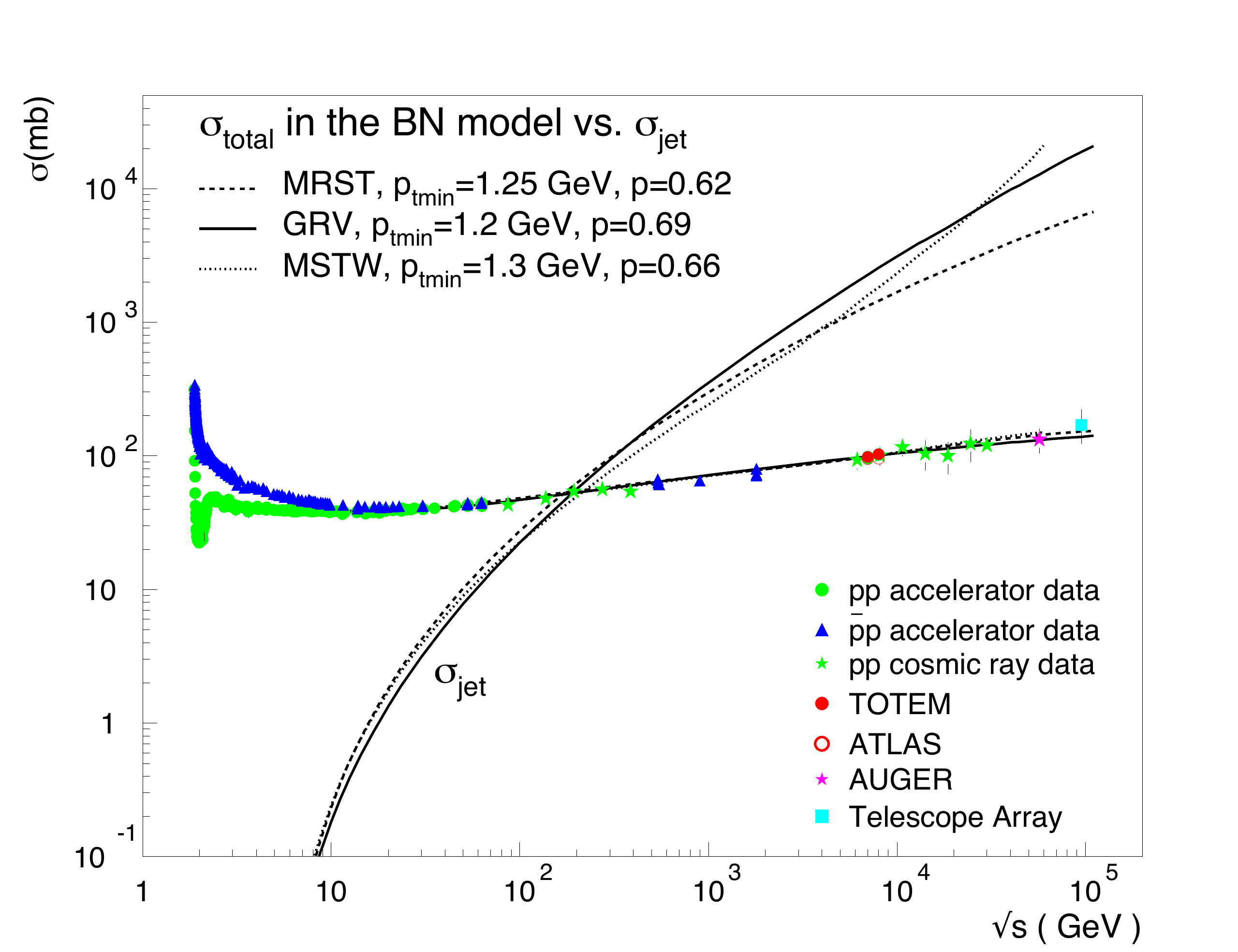}}
\caption{
{The  figure 
shows how the mini-jet cross-section compares with presently available data for $\sigma_{total}$.  The continuous,  dashed and dotted curves correspond to  three different parton density functions, such as   Gl\"uck, Reya and Vogt (GRV) \cite{Gluck:1998xa,Gluck:1991ng}, Martin, Roberts, Stirling and Thorne (MRST) \cite{Martin:1998sq} and Martin, Stirling, Thorne and Watt (MSTW) \cite{Martin:2009iq}. The corresponding curves  over  the total cross-section data are  obtained with the BN model referred to in the text}.}
\label{fig:minijet-vs-total}
\end{figure}

The result of our calculation is shown in Fig. ~\ref{fig:minijet-vs-total} for three different LO PDF sets, 
together with presently available data for the total cross-section \cite{Antchev:2016vpy,Aaboud:2016ijx,Aad:2014dca,
Collaboration:2012wt,Abbasi:2015fdr}.
The comparison between the  energy rise of $\sigma_{jet}(s;p_{tmin})$, [from here on, the terms {\it jet} and {\it mini-jet} are used interchangeably], and  the actual total cross-section 
highlights  the well known fact that, around ISR energies, hard QCD collsions, as calculated to LO,   
start becoming important, but then rising too much.
This difficulty is solved in the BN model  by dressing the mini-jet cross-section with  the phenomenon of soft gluon 
emission which dampens the rise of the  parton-parton cross-sections,   and  embedding them in  
the formalism  of eikonalization, which ensures unitarity. Soft gluon emission in impact parameter 
space then provides the large distance  cut-off which allows satisfaction of the Froissart bound 
\cite{Grau:2009qx}. For 
partial completeness, we shall outline here the  main points of our approach to resummation of  soft gluon emission in hadronic process.

 \subsection{Soft gluon resummation in the infrared region}
 
Together with the PDF driven mini-jet contribution, the core feature of  our working model lies in the 
 impact parameter distribution of the pQCD term,  $A_{BN}(b,s)$, which was  obtained as the 
 Fourier transform of the resummed probability for soft gluon emissions accompanying any QCD 
 scattering process. As we discuss next, the subscript $BN$ refers to our choice of exploiting the 
 full range of soft gluon momenta, down to $k_t=0$, in the spirit of the Bloch and Nordsieck description 
 of soft quanta emission.   
 
 For the resummed soft gluon distribution,   we had proposed \cite{Corsetti:1996wg} to start from the  semi-classical  expression  
 {\cite{Etim:1967aa}}
        \begin{eqnarray}
    \Pi (K_t,s)=
     \int  \frac{d^2{\bf b}}{(2\pi)^2} e^{-i{\bf K}_t \cdot {\bf  b} -h(b,s)} \label{eq:dp2k}\\
    h(b,s)=\int
    d^3 {\bar n}({\bf k}, s)
    [
    1-e^{
    i{\bf  k}_\perp\cdot {\bf  b}
    }
    ]
    \label{eq:hbs}
    \end{eqnarray}
with $d^3 {\bar n}({\bf k})$  being  the  single soft 
{quantum}  spectrum, 
which is exponentiated and regularized through  resummation.
{
Eqs.~(\ref{eq:dp2k}) and  (\ref{eq:hbs}) exhibit a crucial result of the resummation technique developed in \cite{Etim:1967aa}, i.e. the cancellation at semi-classical level of the QED singularities arising from infrared emission and virtual exchanges. Such cancellation follows from imposing energy-momentum conservation to  resummation of  soft quanta emitted  through Poisson distributions, as we outline in  Appendix \ref{app:etp}. }

Unlike $ \Pi (K_t,s)$, which can be obtained through a semiclassical calculation,   the application of the above  technique  to elementary particle processes
requires the spectrum $d^3{\bar n}({\bf k})$ 
 to be determined from quantum field theory, in particular from QCD, in the case of soft gluon emission.

Within the context of  the Bloch-Nordsieck approach,  one can find an early discussion of the probability distribution $ \Pi (K_t,s)$ 
for particle production in strong interactions    with a constant  large coupling in \cite{PancheriSrivastava:1976tm}.  
Applied to Drell-Yan production processes,  the QCD case of running $\alpha_s$ was  examined  in    
 \cite{Dokshitzer:1978yd}  and  \cite{Soper:1980mq} in the Leading Logarithmic Approximation, 
and  by  Parisi and Petronzio (PP)   \cite{Parisi:1979se} within 
 the context of the Bloch-Nordsieck approach.
 In particular, the expression proposed in  \cite{Parisi:1979se} for the function $h(b,s)$ 
 reads:
\begin{equation}
h^{(PP)}(b,s)= \frac{4}{3\pi^2}\int _{M^2}^{Q^2} d^2 {\bf k}_\perp [
    1-e^{i{\bf  k}_\perp\cdot {\bf  b} }]\alpha_s(k_\perp^2)\frac{\ln (Q^2/k_\perp^2)}{k_\perp^2} \label{eq:hpp}
\end{equation}
with a lower limit of integration $M^2\neq 0$ and using the asymptotic freedom expression for $\alpha_s$. 
The contribution of the infrared region, $k_\perp^2\le M^2$  was incorporated in an intrinsic transverse 
momentum factor, with the assumption  that the neglected terms coming from this region would not have a 
singular behavior which could affect the result.

On the other hand, our long held proposal  \cite{Nakamura:1984na,Corsetti:1996wg}
 is to calculate the probability  resummation function $ \Pi (K_t,s)$ down into the infrared region, 
 as relevant to the large $b$-behaviour of the total cross-section, since this is a region where a singular 
 behavior might manifest itself, through a confining potential. Thus, in our approach the single gluon  spectrum  
 depends on  the coupling $\alpha_{IR} (k_t)$ in the infrared region. Our modeling of such behavior has been 
 discussed in many papers, in particular we have a thorough discussion in \cite{Grau:1999em} and \cite{Pancheri:2016yel}.
 
Let $\Lambda$ be an infrared scale separating the asymptotic freedom QCD regime from the non perturbative one,   
then our   phenomenological ansatz for the coupling as $k_t\rightarrow 0$ \cite{Nakamura:1984na,Corsetti:1996wg}, leads to  
 \begin{equation}
 \alpha_{IR}(k_t) \propto [\frac{\Lambda}{k_t}]^{2p} \ \ \ \ \ \ \ \ k_t << \Lambda
 \end{equation}
The above limit  can be justified by a semi-classical argument about confining potentials 
\cite{Nakamura:1984na,Grau:1999em}, 
and the parameter $p$    could be considered as  parametrizing  such complex processes as resummation of multi soft   
gluon couplings. For integrability of the rhs in Eq. (\ref{eq:hbs}) on  the one hand, and for a correspondence to a 
rising potential on the other, the parameter $p$ is limited to the range $1/2<p<1$ \cite{Grau:2009qx}.

With such ansatz for $\alpha_s(k_t\rightarrow 0)$, 
one can calculate the function $h(b,s)$ down into the infrared region. The final calculation of  the normalized 
function $A_{BN}(b,s)$, with the subscript {\it BN}  to indicate the resummation approach we follow,  is done 
by choosing an appropriate value for the singularity parameter $p$ and 
{ specifying  the upper limit of integration in Eq.~(\ref{eq:hbs}), appropriate to the perturbative QCD processes of  
mini-jets. Calling it  ${\qmax}$, it  represents the maximum 
momentum allowed to single gluon emission;  it depends on the energy distribution of the 
emitting partons (hence on the PDFs),
and the perturbative parton-parton cross-section (it was Drell-Yan in \cite{Parisi:1979se}), and  ultimately from    
$\ptmin$.  In our simplified realization of this model,  $\qmax$  is  obtained from   the  expression proposed in   
\cite{Chiappetta:1981bw} as discussed 
in \cite{Grau:1999em}.}


One can then proceed to calculate the average number of hard collisions for  the BN model as
\begin{equation}
 {\bar n}_{hard}(b,s)=A_{BN}(b,s)\sigma^{pp}_{\rm jet} (s;\ptmin)= 
 \frac{
 e^{-h(b,s)}
 }
 {\int d^2{\bf b} e^{-h(b,s)}}\sigma^{pp}_{\rm jet} (s;\ptmin) \label{eq:nbar}
\end{equation}

\begin{figure}
\centering 
 \resizebox{0.8\textwidth}{!}{
 \includegraphics{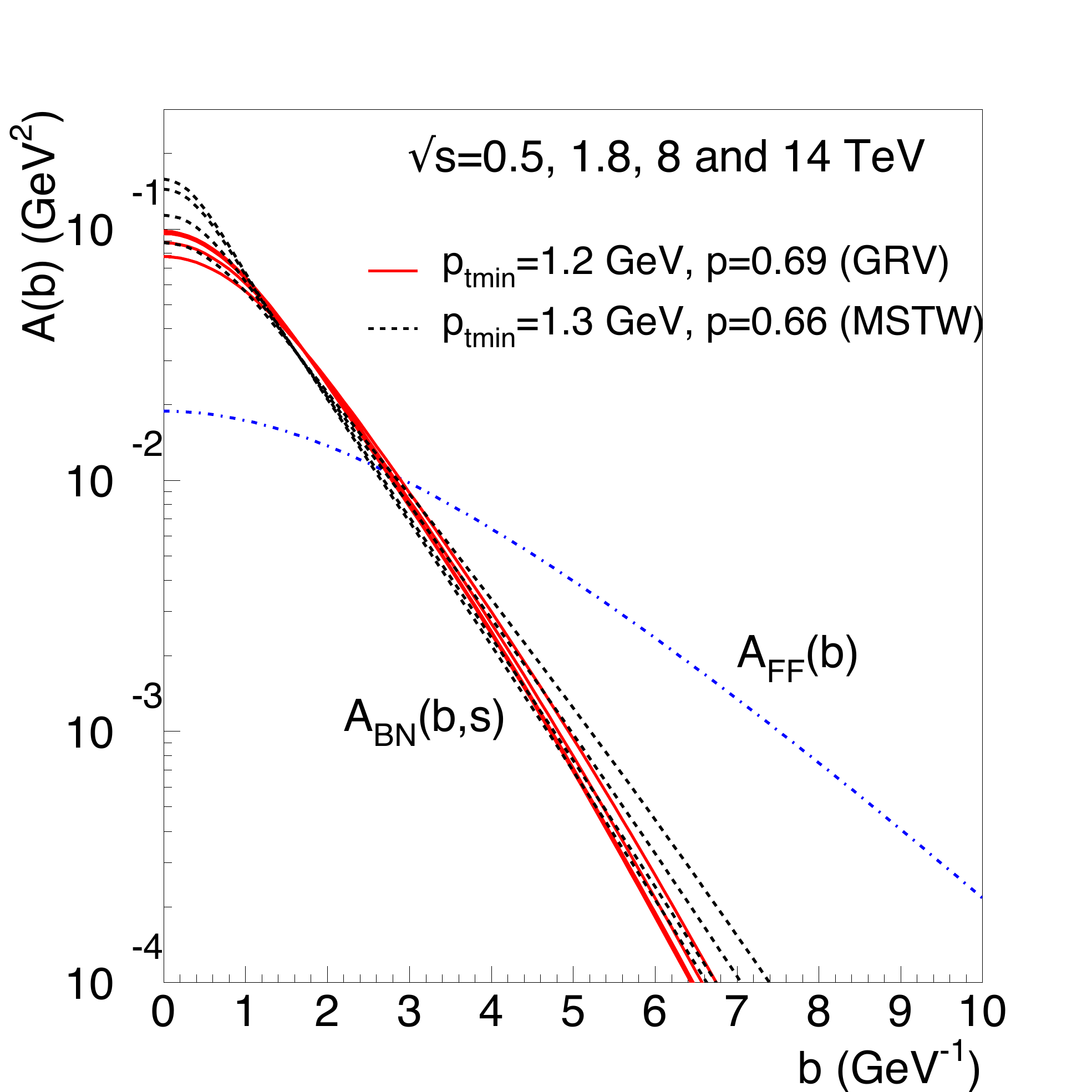}}
 \caption{The dot-dashed line shows the normalized impact parameter distribution  obtained  from 
 the convolution of proton form factors, compared with the  BN model
 for  two different PDFs, and parameters as indicated. 
 { As the energy increases, the distributions in the MSTW case (black dashes) shift more and more towards   $b=0$, whereas GRV curves   (full red) flatten out, with 8 and 14 TeV curves practically indistinguishable. } }
 \label{fig:abn}
 \end{figure}

{In Fig.~\ref{fig:abn} we show the distribution $A_{BN}(b,s) $ for different c.m.energies of the $pp$ system, 
and compare it with an often used impact parameter distribution in total cross-section calculation, namely the convolution 
of proton form factors 
\begin{equation}
A_{FF}(b)=\frac{\mu^2}{96 \pi }(\mu b)^3 K_3 (\mu b)\\
\end{equation}
with $ \mu^2= 0.71$ GeV.  In the figure, two different parametrizations  of the PDFs are used to calculate $q_{max}$ 
(and hence $A_{BN}(b,s)$), MSTW and GRV. The point of interest is two-fold here : for central collisions, i.e. $b\simeq 0$,  
the form factor type distribution (dot-dashed curve)  is much lower than for the mini-jet process, whereas  only a proton form factor type 
distribution survives  at large $b$ values.} 

 \subsection{The total  cross-section in the BN model} \label{sec:total}As 
 well known, and as apparent from Fig.~\ref{fig:minijet-vs-total}, the mini-jet contributions, with their energy dependence, are not sufficient to describe the normalization  of the total cross-section. 
 {Total cross-section data at low energy, i.e. $\sqrt{s}\le 5\div 10$ GeV, suggest to include an additional   contribution}  
which can be   given, in this model,  by the term ${\bar n}_{soft}$ as in Eq. ~(\ref{eq:nbs}), with   $\sigma_{soft}(s)$ parametrized through a best fit to the total cross-section, as
  \begin{equation}
  \sigma_{soft}(s)=48.2 + \frac{101.66}{E_{lab}^{0.99}} -\frac{ 27.89}{E_{lab}^{0.59}} \ \ \ \  (mb)\label{eq:sigsoft}
\end{equation}
  
 The reader would note that in \cite{Godbole:2004kx} a different parametrization of ${\bar n}_{soft}$ had been proposed. 
 We leave to a forthcoming paper a  discussion of these two  different approaches. 
 
 We now see 
  that the calculation of the total cross-section 
 in the BN model, depends on two different sets of parameters:   
 those extracted from  the low-energy regime, with  $\sigma_{soft}$ described by a constant and one or 
 more decreasing powers in energy, and those for the high energy region,
 the latter being:  (i) the choice of the PDFs, (ii) the separation scale between hard and soft processes, 
 $p_{tmin}$, and (iii)  the infrared parameter $p$. The high energy  set characterizes the  energy  
 behavior of the total cross-section as it increases with energy, a behavior driven by QCD mini-jets 
 but regulated by soft gluon emission, modeled by the parameter $p$, as $k_t^{single-gluon}\rightarrow 0$.  

\begin{figure*}
\resizebox{1.0\textwidth}{!}{
\includegraphics{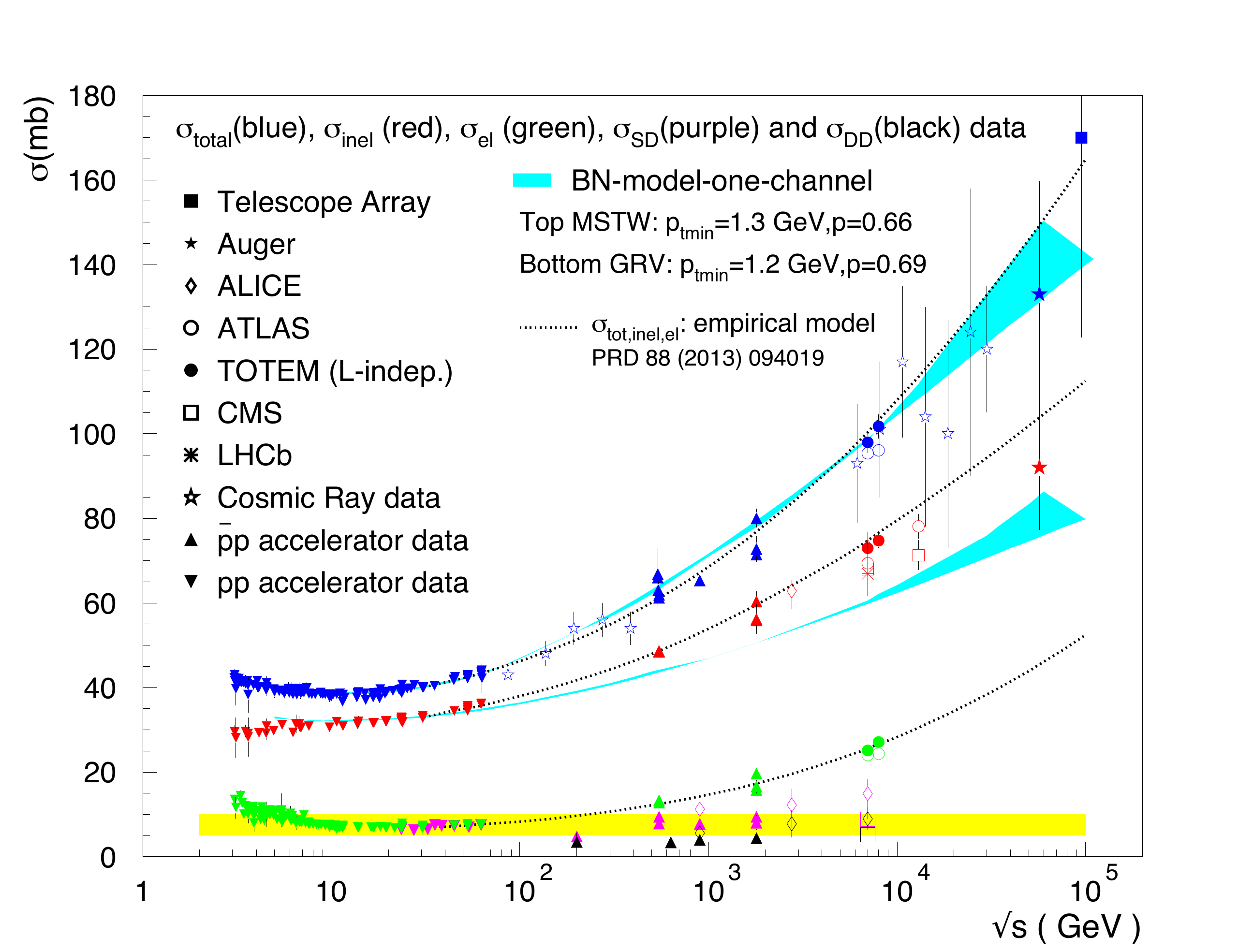}}
\caption{The 
figure  shows the band of expected results for the total and inelastic  cross-section for the BN model for two sets of PDFs, 
and predictions from the empirical model of \cite{Fagundes:2013aja} for total, elastic and inelastic cross-sections. 
For the inelastic cross-section, only data with extrapolation to the full phase space are shown. Other data and 
references for the inelastic cross-section measurements at LHC  are summarized in Table \ref{tab:inel}. 
{For a comparison, diffractive data are also shown, with a constant yellow  band to  guide the eye. }}
\label{fig:alltogether}
\end{figure*}

We also notice, in Fig.~ \ref{fig:minijet-vs-total}, that the 
     different  trend of the mini-jet cross-sections  in the high energy region, due to    the small-x behavior of the 
   parton-parton cross-section from different PDFs, is  much smoothed down in the total cross-section. This is due 
   to the interplay between mini-jet rise and the accompanying  soft gluon emission which dampens it. 
   Such interplay  
   enters through the maximum single gluon momentum $q_{max}$ which is proportional to $p_{tmin}$, the {\it fixed} mini-jet scale. 
   The dependence on densities and $p_{tmin}$ however is not eliminated completely. 
   This appears      clearly in
   Fig. ~\ref{fig:alltogether}, where 
the actual calculation of the total cross-section  from  Eq.~(\ref{eq:sigtot}) is presented in a linear-log scale (rather than log-log as in Fig.~\ref{fig:minijet-vs-total}).

   As discussed and seen in \cite{Fagundes:2015vba},   tuning of  the parameters leads to an optimal description 
   of the total cross-section data up to  $\sqrt{s}=7$ and $8$  TeV, both using ``old" densities, such as GRV, 
   as well as  using more recent parametrizations such as MSTW.  However,   the small-x behavior of the 
   parton-parton cross-section  still leads to $(10\div20)\%$ uncertainties when  extrapolation is done to higher energies 
   such as those reachable through cosmic ray experiments. 
   
   In the next subsection, we shall discuss the other 
   {data and }curves appearing in this figure.

 \subsection{The inelastic cross-sections} \label{sec:inel}
For  estimates  of the survival probabilities \cite{Bjorken:1992er}, the quantity of interest is the inelastic cross-section in impact parameter space.
{In single  channel models 
 this can be  obtained through the elastic amplitude}
\begin{equation}
 {\cal F}(s,t)= i \ \int  b db J_o(qb)[1-e^{-\chi_I (b,s)}  ] \label{eq:BNampli}\\
\end{equation}
with $q^2=-t$,
 namely 
    from the equation
 \begin{equation}
  \sigma_{inelastic}= \int d^2{\bf b} [1-e^{-2\chi_I (b,s) } ]\equiv   \int d^2{\bf b} [1-P_{no-inel}(b,s)]\label{eq:siginel}
 \end{equation}
 with 
 \begin{equation}
 P_{no-inel}(b,s)=e^{-{\bar n}(b,s)}\equiv e^{-{\bar n}_{soft}(b,s)-{\bar n}_{hard}(b,s)}\label{eq:pnoinel}
 \end{equation}
 {in a two-component eikonal as described before}. However  one problem arises: as discussed in \cite{Fagundes:2015vba} and clearly seen in Fig.~\ref{fig:alltogether} the  inelastic cross-section obtained  from Eqs.~(\ref{eq:siginel}) and (\ref{eq:pnoinel}), and estimated with the parameters leading to the good description of  $\sigma_{total}$,  reproduces LHC inelastic data only  in a limited range, $\xi =M^2_X/s  \gtrsim 5 \times 10^{-6}$
 {falling short of  the full phase space extrapolated data.}
 
  {A model independent  estimate of the inelastic cross-section is shown by the dotted lines in Fig.~\ref{fig:alltogether}. This estimate is obtained    as $\sigma_{inel}=\sigma_{tot}^{emp}-\sigma_{el}^{emp}$ by mean of an empirical parametrization of all the differential $pp$ cross-section data from ISR to LHC, based on  the elastic amplitude
   \begin{equation}
    {\cal A}(s,t)=i[F^2_P(t/t_0)\sqrt{A(s)}e^{B(s)t/2}+e^{i\phi(s)}\sqrt{C(s)}e^{D(s)t/2}] \label{eq:mpb2}
   \end{equation}
  where $   F^2_P(t)$ is the square of the proton form factor, i.e. $F_{P}(t/t_0)=1/[(1+|t|/t_0)]^2$. Details of this model, which is a modified version of the 1974  Phillips and Barger proposal \cite{Phillips:1974vt}, 
  can be found in \cite{Fagundes:2013aja} and are reproduced here  in the Appendix \ref{app:empirical}. 
  { For a new version of the model, now augmented to describe the $\rho$ parameter,  see \cite{RuizArriola:2016ihz}. }

{Thus, the empirical model applied to the inelastic cross-section confirms the extrapolations to the full phase space as obtained through MC simulations or other models.
At the same time,   the single channel two-component BN model, we just described, has so far not included 
{energy dependent }diffraction processes. Data for these type of events are displayed in Fig.~\ref{fig:alltogether}, respectively from ISR \cite{Armitage:1981zp}, UA5  \cite{Ansorge:1986xq}, UA4  \cite{Bernard:1986yh}, CDF\cite{Abe:1993wu}, \cite{Abelev:2012sea} and  CMS\cite{Khachatryan:2015gka}.}

{One  obvious missing element in the single channel model we have proposed is  Single Diffraction. As seen from Fig.~\ref{fig:alltogether}, this process, unlike Double Diffraction, shows    an energy dependence  characteristic of QCD processes, namely its contribution increases with energy. Indeed, while the BNmodel so far includes QCD processes such   as gluon-gluon collisions and the accompanying resummed soft gluon emission,  it misses one more process  which can give energy dependence to the cross-section through perturbative QCD, namely  hard gluon bremsstrahlung from the proton and its accompanying soft gluon emission, as well.  This process, at the origin of Single Diffraction contributions, is correlated to the emitting proton and its inclusion in a single-channel model has so far been difficult.}

  However, 
  lacking a clear understanding of diffraction   in mini-jet type models,     we propose that   the quantity 
  $P_{no-inel}$ thus calculated 
 can  be used  to estimate survival probabilities when Single Diffractive events are not excluded, and proceed to do so in the next section.

{We conclude this section with a comparison of our  single channel BN model with recent  experimental results for the inelastic cross-section, in the measured phase regions, as   shown in Table \ref{tab:inel}.}
\begin{table*}
\caption{
The inelastic cross-section at LHC obtained from the single channel two-component BN model, and its comparison with existing data at LHC and cosmic ray energies. %
{
The last column shows  the estimate obtained (by subtraction)  from the empirical model  of \cite{Fagundes:2013aja}.
}
}\label{tab:inel}
\scriptsize{
\begin{tabular}{|c|c|c|c|c|c|c|}
\hline
$\sqrt{s}$& $\sigma_{inel}$                                                          &Kinematic range                          &Experiment                         & $\sigma_{inel}^{BN}$&Emp.\\
TeV        &  mb                                                                              & $\xi_X =M^2_X/s$                      &                                           & mb                                          & mb \\
               &                                                                                      &and $\xi_Y =M^2_Y/s$               &                                           &GRV-MSTW                           & \cite{Fagundes:2013aja}\\
\hline \hline
2.76       & $62.8  ^{+2.7} _{-4.2}$                                       &    full- sim.                                   & ALICE \cite{Abelev:2012sea}  &                                       & \\
\hline
7            &                                                                                         &  no SD                                        &                                                       &59.8-60.5 & \\
           & $60.3\pm 0.5(syst)\pm 2.1(lum)$                             &  $\xi_X >5\times 10^{-6}$                      & ATLAS \cite{Aad:2011eu}             &         & \\
              & $60.2\pm 0.2 (stat) \pm 1.1 (sys) \pm 2.4 (lum)$          &   $\xi_X >5\times 10^{-6}$                    &CMS  \cite{Chatrchyan:2012nj}  &                             & \\
              & $62.1^{+1.0}_{-0.9}(sys)\pm 2.2 (lum)$                          &$\xi_X >5\times 10^{-6}$                       & ALICE \cite{Abelev:2012sea}  &                               & \\  
              & $55.0\pm 2.4 $   & $p_T>0.2 GeV/c, 2.0<\eta<4.5$                                                               &LHCb \cite{Aaij:2014vfa}              & &\\  
              &$71.34 \pm 0.36(stat) \pm 0.83(syst)$                    &   full-by subtraction                       & ATLAS \cite{Aad:2014dca}            & & 74.8\\                                                   
              & $72.9 \pm1.5$                                                             &    lum-independent -full                 &TOTEM \cite{Antchev:2013iaa}     & &\\
              & $68.0 \pm 4.0 (model) \pm 2.0(sys)\pm 2.4 (lum)   $&  full-MC simu; &CMS   \cite{Zsigmond:2012vc}   \footnote{also in CMS-PAS-FWD-11-001, 
                superseded by  \cite{Chatrchyan:2012nj}.}   &    & \\

              & $66.9\pm 2.9 (exp)\pm 4.4 (extr)$ & full-Pythia 6                                                                                      &LHCb \cite {Aaij:2014vfa}      &       & \\

              &$73.2^{+2.0}_{-4.6}(model) \pm 2.6 (lum)$                        & full  -diff model                              & ALICE \cite{Abelev:2012sea}&    & \\
 \hline
 8            &      & no SD                                     &  &60.7-62.1 & \\
            & $ 74.7\pm 1.7   $                                                      & full-MC   simul.                                      & TOTEM \cite{Antchev:2013paa} & & \\
                            & $71.73\pm 0.15 (stat)\pm 0.69 (sys)$         & full-by subtraction                                 & ATLAS  \cite{Aaboud:2016ijx} & &76.6 \\
              \hline
\hline
13           &                                                                                         &                                     no SD                    &             & 64.3- 66.6                         & \\                                   
          &$65.77\pm 0.03(stat)\pm 0.76(sys)\pm 1.78 (lum)$& HF $\xi >10^{-6}$                                  & CMS \cite{VanHaevermaet:2016gnh} & &  \\
      &   $   68.1 \pm 0.6(exp) \pm 1.3(lum)                                                                                 $  & $  \xi > 10^{-6}$         &ATLAS \cite{Aaboud:2016mmw} &  & \\
&$66.85\pm 0.06 (stat)\pm 0.44 (sys)\pm 1.96 (lum)$&HF+CASTOR $\xi_X >10^{-7}, \xi_Y>10^{-6}$&CMS \cite{VanHaevermaet:2016gnh}& & \\
           &  $71.26 \pm0.06 (stat) \pm 0.47 (sys) \pm 2.09 (lum) \pm 2.72 (ext)$ & extr. all models          & CMS \cite{VanHaevermaet:2016gnh} & & \\
                & $ 78.1 \pm 0.6(exp)\pm 1.3(lum) \pm 2.6 (ext)                                                                                   $& extr. - full                     &ATLAS \cite{Aaboud:2016mmw} & & 82.9\\
                \hline
14      &                                                                                                                      &      No SD                                 &                                                           &64.8- 67.4 &  \\   
14      &                                                                                                                      &                                                   &                                                           &                   &  83.9\\             
                \hline
   57                                                                                               &                                                & No SD                                                    &                                                                 &75.6-85.4& \\
$57$   \footnote{with error  $\pm 0.3 (stat) \pm 6 (sys)$.}  &$92 \pm ^{13.4}_{14.8}$     & full-from Glauber and other effects   & AUGER \cite{Collaboration:2012wt} &   & 103.8 \\
              \hline \hline
\end{tabular}}
\end{table*}

\section{ Survival Probabilities}
Let us recall  early discussions  of survival probability \cite{Dokshitzer:1991he,Bjorken:1992er}
that arose in considerations of 
a hadronic collision at an impact parameter $b$ producing a final state  characterized by  
energy  scales much larger than  those  of the  soft and semi-hard background of hadronic collision. 
Such a final state  can be  high $p_t$ jet pair  production, or  Higgs production, for instance,
and, we look for events with {\it no}   hadronic activity in the central region.

 Let $ S^2(b,s)$ be  the distribution for observing one such high $p_t$ process with cross-section
 $\sigma_{hard-scale}(s)$ and no additional inelastic collisions  \cite{Bjorken:1992er}. A simplified factorized model for such distribution can be written as
    \begin{equation}
   S^2(b,s)=  \sigma_{hard-scale}^{AB}(s)  {\cal H}(b,s)P_{no-collisions}(b,s) \label{eq:survgeneral}
    \end{equation}
where ${\cal H}(b,s)$ is the  distribution  in impact parameter space of those partons participating to the collision leading 
to the  production of  H (the hard-scale process).
Then the distribution $S^2(b,s)$  can be  integrated and normalized
and the  average survival probability distribution is obtained  from the simplified expression
 \begin{equation}
{\cal S}^2(s)\equiv<| S(b,s)|^2>=\int d^2{\bf b} A(b,s)e^{-{\bar n}(b,s)}\label{eq:survival} 
\end{equation}
having used Eq.~(\ref{eq:pnoinel}) and with 
\begin{equation}
A(b,s)=\frac{{\cal H}(b,s)}{\int d^2 {\bf b} {\cal H}(b,s)}
\end{equation}
Leaving aside for the time being the question of the missing piece of the inelastic cross-section
in  single channel eikonal minjet models such as the one  described earlier, we can write
\begin{equation}
P_{no-collisions}(b,s)=
 P_{no-soft-collisions}(b,s) P_{no-mini-jets}(b,s) \label{eq:pnocoll}
\end{equation}
where the first factor on the r.h.s.  excludes the presence of 
{soft partons from} events for  which the  cross-section  
is either constant or decreasing.
 This term alone  does not exclude production of mini-jets. Instead, these processes, 
which can be described by perturbative QCD, as we have seen, and constitute  the hadronic background for which partons 
exit the collision with  $p_t>p_{tmin}\simeq 1\ GeV$ accompanied by   the infrared initial state emission, are  suppressed through the factor  
$P_{no-mini-jets}(b,s)=exp[-{\bar n}_{mini-jets}(b,s)]$.

 In cases where one puts a $p_t$-cut (say, 1 GeV) 
to eliminate the mini-jet   emission 
 (as when  the  hard process to select is production of a color singlet, for instance),  one would have to consider 
 \begin{equation}
  S^2(b,s)=  \sigma_{hard-scale}^{AB}(s) {\cal H}^{mini-jets}(b,s) P_{no-mini-jets}(b,s) 
    \end{equation}
    but notice that not all  low $p_t$ activity is excluded, since  some hadronic activity from 
    ${\bar n}_{soft}(b,s)$ has not been excluded.
    
 If absence of both soft collisions and mini-jets is required, then one should use  the full probability  
 $P_{no-collisions}(b,s) $ as in Eq.~(\ref{eq:survgeneral}). We shall now address   
 the question as to which impact parameter distribution is appropriate to a given measurement. 
 In what follows, we shall see what is involved in such calculations and compare with existing model predictions.
     
\subsection{SP results
from 
BN-2008 and 
Block {\it et al.}-2015 estimate from QCD inspired models 
}

Let us start with SP estimates from the BN model, as done originally  in 2008 \cite{Achilli:2007pn} 
and  revisit it  
in order to compare with (2015) results from Block and collaborators \cite{Block:2015sea}. For quark initiated processes, 
the total survival probability of the gap in this QCD inspired model  is obtained from    the expression
\begin{equation}
<| S(b)|^2>=\int d^2{\bf b} A(b,\mu_{qq})e^{-2\chi_I(b,s)}\label{eq:survBlock}
\end{equation} 
with $A(b,\mu_{qq})$ the distribution of quarks in the proton, for which the relative  parton-parton cross-section 
is decreasing. In this single  channel eikonal model \cite{Block:1991yw,Block:1998hu},
$\chi_I(b,s)$ is obtained as the contribution from three terms: gluon-gluon, quark-gluon and quark-quark collisions, i.e.
\begin{equation}
\chi_I(b,s)=W(\mu_{gg}) \sigma_{gg}(s) + W(\mu_{qg}) \sigma_{qg}(s)+W(\mu_{qq}) \sigma_{qq}(s)
\end{equation}
with $W(\mu)$ obtained from a convolution of dipoles, and different scales $\mu_{ij}$ in correspondence 
with three basic cross-sections $\sigma_{ij}$, with different energy behaviors. These parameters were tuned to 
the large set of elastic and total cross-section data available before the LHC operation.
As we shall see in more detail later, this model predicts  rather large survival probabilities at LHC when  
compared with recent estimates from the Durham-St. Petersburg group, Khoze, Martin and Ryskin (KMR) 
in  \cite{Khoze:2013dha} and  those from  the Telaviv group of Gotsman, Levin and Maor (GLM)  in \cite{Gotsman:2015aga}.
In the following,  we shall attempt to understand this difference.

 Our earlier estimate of survival probabilities \cite{Achilli:2007pn} 
 was similar 
 {to a previous one by  Block and Halzen \cite{Block:2001ru}}, but we now believe that such estimates should be reconsidered.   
 To understand why (and how), we notice that in \cite{Achilli:2007pn} our estimate was done using
\begin{equation}
<| S(b)|^2>=\int d^2{\bf b} A_{soft}(b,s)e^{-2\chi_I(b,s)}\label{eq:survourold}
\end{equation}
with  $A_{soft}(b,s)$
obtained as  the distribution of partons with final $p_t<p_{tmin}$ in correspondance with non mini-jet collisions. 
{Following our present parametrization of  ${\bar n}_{soft}(b,s)$, we now evaluate Eq.~(\ref{eq:survourold}) using the convolution of proton form factors as discussed in the previous section, 
and the updated  parametrization for $\chi_I(b,s)$ which led to the curves for the total cross-section  
in Fig.~\ref{fig:alltogether}. With this procedure, which we can call    the BN-2008 model for the survival probability}, we    show in Fig.~\ref{fig:sursoft} the  agreement between  
the recent  Block and collaborators results and the estimate from Eq.~(\ref{eq:survourold}).

The reason for the approximate agreement between our calculation and the recent Block 
{\it et al.} result,  lies   in the very similar role played by 
the two distributions  $A(b,\mu_{qq})$   and $A_{soft}(b)$ entering Eqs.~(\ref{eq:survBlock}) 
and (\ref{eq:survourold}): they both   correspond to  parton processes whose cross-section  is not rising with energy. 
At the same time,  in both models, $P_{no-inel}$ is constructed with contributions  from both rising and  constant 
parton cross sections,  with an eikonal such as to reproduce the total cross-section. In the BN model,   
the decomposition of collisions  corresponds to  two types of soft hadronic activity, one coming from processes 
in which the  production of soft partons is energy independent or decreasing,   and one with mini-jet production 
[dressed with infra-red gluons whose number is increasing with energy]
that drives the rise of the  total cross-section. In a similar way, the Aspen  model,  used for the estimate in \cite{Block:2015sea}, 
includes 
three types of contributions, with $gg$ and $gq$  rising with energy, and $qq$ constant or decreasing.  

However, this way to estimate the survival probability  
certainly needs revision for the following reason.
To exclude all hadronic background processes which at high energy show an increase with energy,
 one needs 
 to take into account the  rising contribution from mini-jets or semi-hard collisions from partons 
 whose $b$-distribution is  very different, as shown in Fig.~\ref{fig:abn}
  for the BN model.

 These estimates are 
also compared in Fig.~\ref{fig:sursoft} with the result by Bjorken at $\sqrt{s}=40$  TeV \cite{Bjorken:1992er}, 
where the lowest value  was obtained with a multiplicative model (red line), as we summarize below.

Considering only independent collisions, and an expression as in Eq.~(\ref{eq:survBlock}),  
Bjorken estimated the survival probability to be about 10\%, with numerical estimates from \cite{Block:1991yw}, 
and under the assumption of uncorrelated parton distributions. 

However, when Bjorken included the possibility of hadronic activity clustered around the valence quarks,  
he suggested instead 
the following: 
\begin{equation}
<|S|^2> _{Bj}\simeq 
\frac{
\int d^2{\vec B}
 F(B)
  |S_{pp}(B)|^2
  \int d^2
  {\vec b }
   \ \sigma_{qq}^{Hard} |S_{qq}(b)|^2}
{\int \ d^2{\vec B} F(B) \int d^2{\vec b   \ \sigma_{qq}^{Hard} (b)}
}
=<|S|^2>_{pp} <|S|^2>_{qq} \label{eq:bjorken}
\end{equation}
where $ <|S|^2>_{pp}$ is the survival probability estimated before, whereas $ <|S|^2>_{qq}$ 
is an extra factor. 
The additional term could exclude collisions rising with energy 
and hadronic activity correlated with the valence quarks alone. 
In any case, an additional diminution of the survival probability was expected and a value of 
5\% was  considered more likely (red dot in the figure),  
with {\it a factor 3 uncertainty in either direction.} This is what we have shown in the figure.

{ A comparison is also shown with the LO result by the CMS collaboration \cite{Chatrchyan:2012vc} for the survival probability   in the measurement of  the  diffractive contribution to dijet  production at $\sqrt{s}=7\ TeV$. CMS gives  an estimate of  $S^2 = 0.12 \pm 0.05$ at LO, and a lower value of $S^2=0.08\pm 0.04$  at NLO. A similar more recent (2015)  measurement by the ATLAS collaboration \cite{Aad:2015xis}, not shown in the plot,  uses an estimate of  $S^2=0.16 \pm 0.04 (stat)\pm 0.08 (exp.syst) $  for  dijet production in $\sqrt{s}=$ 7\ TeV $pp$ collisions with large rapidity gaps, 
                         this estimate being  considered to be also consistent with a  central value of $ 0.15$.}  
\begin{figure}[htb]
\centering
\resizebox{0.8\textwidth}{!}{
\includegraphics{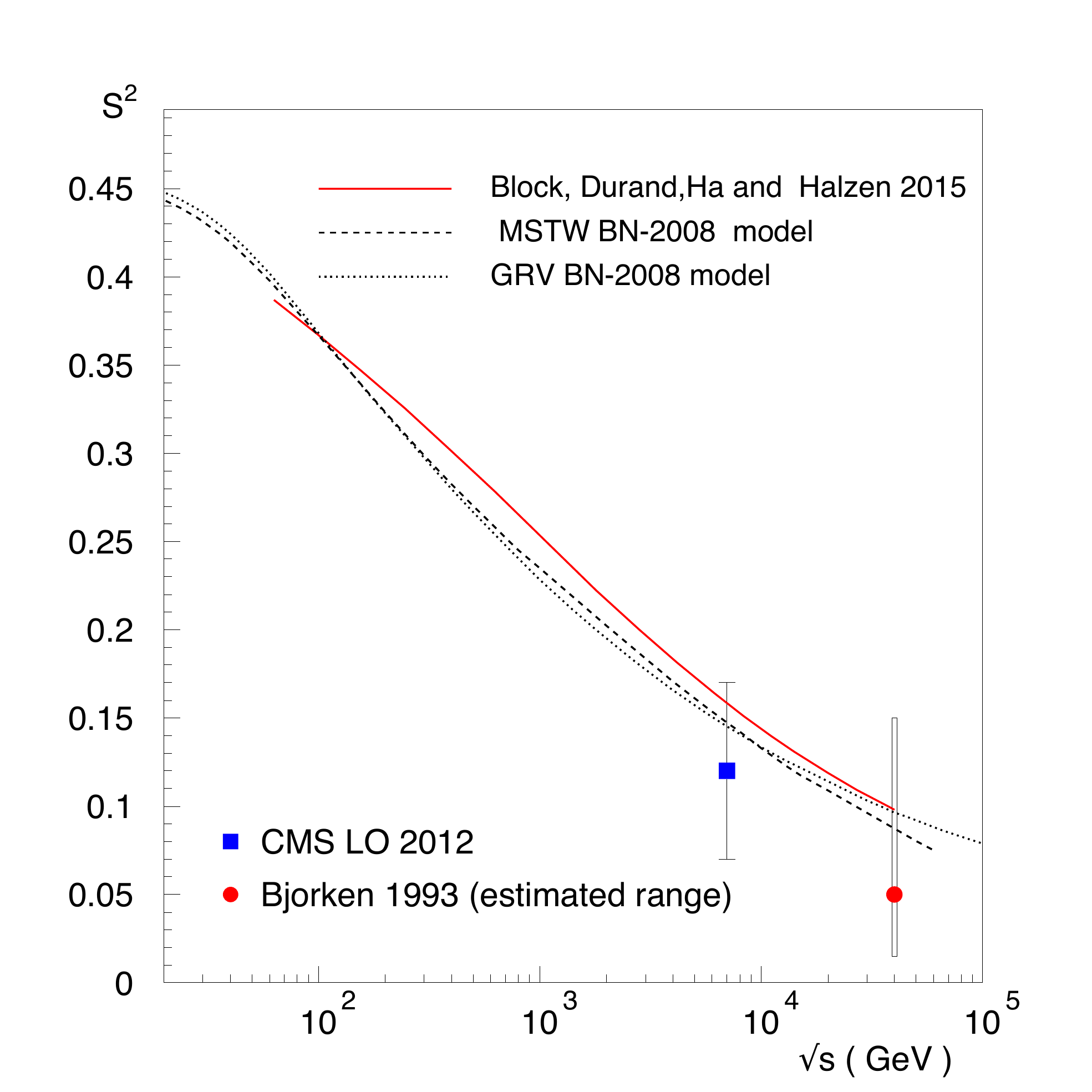}
}
\caption{We show the survival probabilities obtained with the soft process impact parameter distributions  from \cite{Block:2015sea} (full line) and the BN model in \cite{Achilli:2007pn},  compared with the  Bjorken's estimate at 40 TeV \cite{Bjorken:1992er}, based on an  impact parameter  using   a {\it soft}  distribution  first and then  a multiplicative model for hard processes. 
We also show the LO estimate by CMS \cite{Chatrchyan:2012vc}.
}
\label{fig:sursoft}
\end{figure}

In the next subsection, we shall present a different proposal, in which 
 first a split is made between {\it soft} and {\it hard} contributions 
 and then, the fractioned (lack of) hadronic activity from each is summed to construct the SP.   
 We shall 
 compare -with plots and tables- our calculation with the Telaviv and Durham-St Petersburg  models, 
 labelled here, for short, as   GLM and KMR.

\subsection{Our proposal 
with all order resummation of soft gluons}
Let us approach the calculation of the survival probability  in a single channel two-component eikonal model 
such as our BN model, in which one splits the eikonal into a component rising with energy, and 
another component either constant or decreasing. 

To exclude all hadronic uncorrelated activity, one can  distinguish  between  $soft$ and $hard$ collisions as 
participating with different weights to the survival probability,  
\begin{equation}w_{soft/hard}(s)\equiv \frac{\sigma_{soft/jet}(s)}{\sigma_{soft}(s)+
\sigma_{jet}(s)}\equiv \frac{\sigma_{soft/jet}(s)}{\sigma_B(s)}
\end{equation}
with $\sigma_B$ to represent the ``Born term" of the total cross-section,  $\sigma_{jet}$  
being  obtained from Eq.~(\ref{eq:sigjet}) and $\sigma_{soft}$ from Eq.~(\ref{eq:sigsoft}). As one can see 
from Fig.~\ref{fig:minijet-vs-total} , at low energy, $w_{soft} >> w_{hard}$, while their roles 
are exchanged at high energy. 
 Then the contribution to the survival probability will depend on  the relative weights as follows:
\begin{itemize}
\item in the case of emission coming from processes with a cross-section not rising with energy and   
final hadrons with  $p_t<p_{tmin}$, in our phenomenological approach 
\begin{itemize}
\item (i) the $b-$distribution is given by $A_{FF}(b)$, namely follows the form factor distribution, 
with no extra energy dependence,
\item (ii) the probability of no such emission is given by $e^{-{\bar n}_{soft}}$,
\item (iii) the survival probability  is obtained as   
$$<| S(b)|^2>_{soft}=
\int d^2{\bf b} A_{FF}(b,s)e^{-{\bar n}_{soft}(b,s)} $$ 
\end{itemize}
\item in the case of QCD mini-jet processes, for which final hadrons have $p_t>p_{tmin}$ and an increasing rising cross-section, 
the b-distribution  is obtained through  soft gluon emission 
accompanying the mini-jet collision, and 
$$<| S(b)|^2>_{hard}= 
\int d^2{\bf b} A_{BN}(b,s)
e^{-
{\bar n}_{hard}(b,s) 
}
$$
\end{itemize} 
Our  proposal is that the survival probability -to exclude  hadronic activity in the central region- is given 
by
\begin{equation}
\bar{\mathcal{S}}^{2}_{total} (s)= \bar{\mathcal{S}}^{2} _{soft}(s)+\bar{\mathcal{S}}^{2}_{hard} (s)
\equiv w_{soft}(s)<| S(b)|^2>_{soft}+w_{hard}(s)<| S(b)|^2>_{hard} 
\label{eq:survsum}
\end{equation}
With the {\it caveat} that  
{diffractive  events  are either poorly or not at all   
described by the single channel model} and hence  are not} 
{excluded by } 
$P_{no-inel}$, 
we now proceed to calculate the survival probabilities and 
compare it with other models.

\begin{figure}
\resizebox{0.8\textwidth}{!}{
\includegraphics{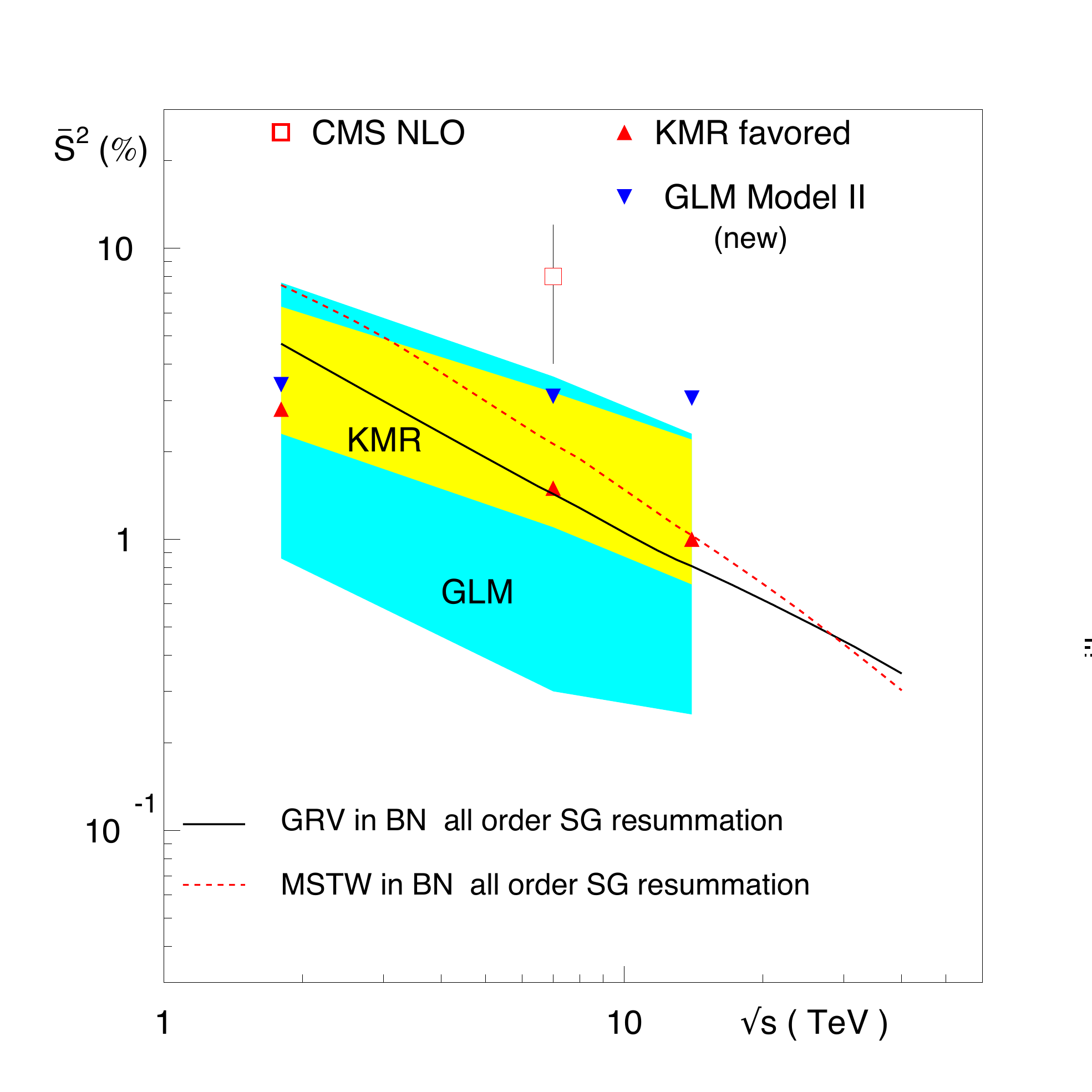}}
\caption{
The (black and red) curves
   indicate the estimated Survival Probabily Rapidity Gaps (in percentage) in the LHC region, in the additive model we propose in Eq.~(\ref{eq:survsum}), using impact parameter distributions obtained from the BN model for the mini-jet component, with MSTW (dashes) or GRV (full) LO PDFs, with all order Soft Gluon (SG) resummation. Comparison with  ranges estimated in    the models by  GLM (cyan band) and KMR (yellow band),   and with  LHC measurements at 7 TeV by CMS at NLO \cite{Chatrchyan:2012vc} 
    is also shown.}
\label{fig:surv}
\end{figure}

We present the results of our proposal 
in Table \ref{tab:GRV-MSTW} for the two types of densities used to describe the inelastic (and hence the total) cross-section,
\begin{table}
\caption{Survival probabilities, soft, hard and total,  in the TeV region, in the additive model we propose, using    MSTW densities ($p_{tmin}=1.3$ GeV and $p=0.66$), and GRV ($p_{tmin}=1.2$ GeV and $p=0.69$). All values are given in percentages,  with the values taken by   $ \bar{\mathcal{ S}}^{2} _{total}$   plotted Fig.~\ref{fig:surv}.}
\label{tab:GRV-MSTW}
\begin{tabular}{|c|c|c||c|c||c|c|}
\hline
$\sqrt{s}$  &\multicolumn{2}{|c|}{$ \bar{\mathcal{S}}^{2} _{soft}$}&\multicolumn{2}{|c|}{$ \bar{\mathcal{S}}^{2} _{hard}$}&\multicolumn{2}{|c}{$ \bar{\mathcal{S}}^{2} _{total}$} \\
TeV & GRV  & MSTW         & GRV  & MSTW            & GRV  & MSTW   \\ 
\hline \hline 
1.8 & 3.17 & 4.77 & 1.53 & 2.70 & 4.70 & 7.47 \\ 
\hline 
2.76  & 2.15 & 3.38 & 1.06  & 1.95 & 3.21 & 5.33 \\ 
\hline
7.0  & 0.942 & 1.32 & 0.490  & 0.810 & 1.43 & 2.13 \\
\hline
8.0  & 0.839 & 1.17 & 0.440 & 0.722 &  1.28 & 1.89 \\
\hline 
13  &  0.554 & 0.681 & 0.297 & 0.433 & 0.851 & 1.11 \\ 
\hline 
14  &  0.526 & 0.669 & 0.282 & 0.425 & 0.808 & 1.03 \\ 
\hline 
40  &  0.222 & 0.182 &  0.124 & 0.121 & 0.346 & 0.303 \\ 
\hline \hline 
\end{tabular} 
\end{table}
{and  show in Fig.~\ref{fig:surv}  the values taken   by $ \bar{\mathcal{S}}^{2} _{total}$     for GRV and MSTW densities, in comparison with GLM and KMR estimated ranges. }
We also show comparison with the NLO CMS estimate {\cite{Chatrchyan:2012vc}. Since our present proposal is obtained by resummation of soft gluons to all orders, the comparison with NLO result is the  appropriate one. Please notice the change in scales.} 
  
\begin{figure*}
\resizebox{1.0\textwidth}{!}{
\includegraphics{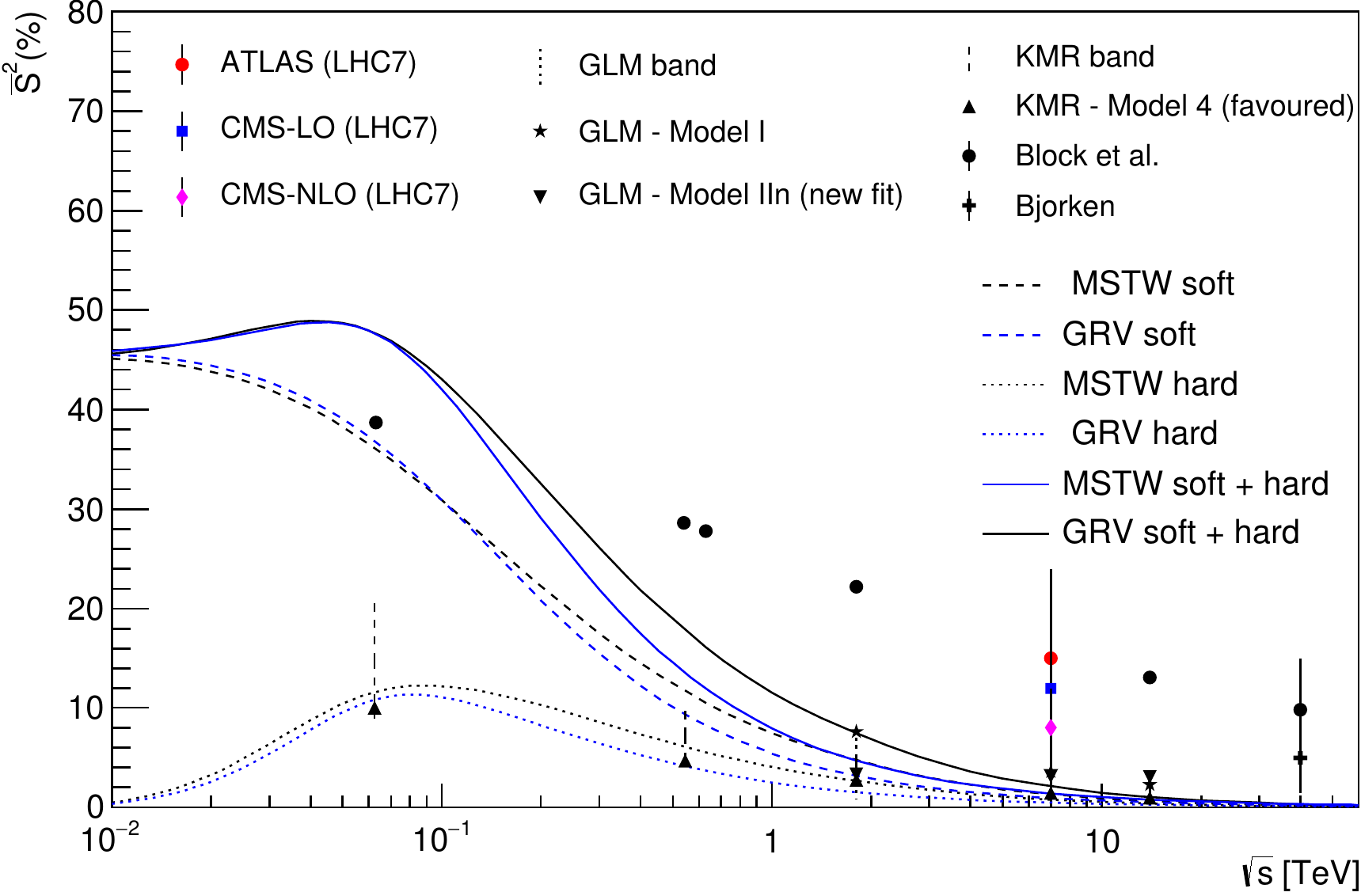}}
\caption{
\label{fig:what}
Comparison of the predictions for the survival probability gaps estimates  (in percentage) in the context of the BN model,   with other models described in the text and with CMS \cite{Chatrchyan:2012vc} and ATLAS \cite{Aad:2015xis} measurements at LHC7.}
\end{figure*}

These results are  now summarized  in Table \ref{tab:comparisons}, and Fig.~\ref{fig:what}, 
{where we  compare  our proposal with   the experimental estimates by ATLAS and CMS collaborations, with those by  Block, Durand, Ha and Halzen, 
  the two Reggon-Pomeron models we have mentioned,  and the  40 TeV  range of values estimated by Bjorken \cite{Bjorken:1992er}.
   For the BN model,  we   also show the separate estimates for $\bar{\mathcal{S}}^{2} _{soft}$ and $ \bar{\mathcal{S}}^{2} _{hard}$.}
\begin{table}
\centering
\caption{Survival probability predictions of the models by Block-Durand-Ha-Halzen (BDHH) \cite{Block:2015sea}, Khoze-Martin-Ryskin (KMR) \cite{Khoze:2013dha} and Gotsman-Levin-Maor (GLM) \cite{Gotsman:2015aga}, and the range of prediction by Bjorken (BJ) \cite{Bjorken:1992er}. All values are given in percentages.The BN model range includes calculation with two different PDFs.}
\label{tab:comparisons}
\vspace*{.2cm}
\scriptsize{\begin{tabular}{|c|c|c|c|c|c|}
\hline \hline 
$\sqrt{s}$  & $\bar{\mathcal{S}}^{2}_{BDHH}$  & $\bar{\mathcal{S}}^{2}_{KMR}$ & $\bar{\mathcal{S}}^{2}_{GLM}$ &$\bar{\mathcal{S}}^2_{BN}(I)$&$\bar{\mathcal{S}}^2_{BJ}$\\ 
  TeV          &                                  &                              &                              & GRV-MSTW &  \\
\hline 
0.063 & 38.7 $\pm $ 0.6 & $8.7 - 20.8\ (10)$ & $-$ & &  \\ 
\hline 
0.546 & 28.6 $\pm $ 0.5 & $4.1 - 10.3\ (4.7)$ & $-$ & &  \\ 
\hline 
0.630 & 27.8 $\pm $ 0.5 & $-$ & $-$ & &  \\ 
\hline 
1.8 & 22.2 $\pm $ 0.5 & $2.3 - 6.3\ (2.8)$ & $0.86  - 7.6\ (3.34)$ &4.70 - 7.47&    \\ 
\hline 
7.0 & $-$ & $1.1 - 3.2\ (1.5)$ & $0.3 - 3.63\ (3.1)$ &1.43-2.13&   \\ 
\hline 
13 &          &                              &                                    &         0.851-1.11          &  \\
\hline
14 & 13.1 $\pm $ 0.3 & $0.7 - 2.2\ (1.0)$ & $0.25 - 2.3\ (3.05)$ & 0.808-1.03 &   \\ 
\hline 
40 & 9.8 $\pm $ 0.2 & $-$ & $-$ & 0.346 - 0.303 & 1.5-15 (5)  \\ 
\hline \hline 
\end{tabular} }
\end{table}

As already discussed  our estimate is close to Block's only  when the soft distribution, which we take to 
be the folding of two proton form  factors,  is used. 
{On the other hand, we find    our results to be consistent  with  the range of values 
coming from different models by the Durham  \cite{Khoze:2013dha} and  Telaviv group \cite{Gotsman:2015aga}.}

In the table,  the Durham model results  are obtained using the GW formalism in a two channel eikonal  model, 
which includes low mass diffractive dissociation, and is able to predict both elastic and diffractive cross-sections. 
Four different models are discussed, all of which give similar good fit to  the various cross-sections, but  have   
different values for $|S|^2$. The difference  is ascribed to depend on  the details of the Good and Walker splitting 
and  hence to the impact parameter density of the GW states.  The authors' favored model  is indicated between 
parentheses and corresponds to energy dependent coupling of the  triple Pomeron.  

The table also displays a band of prediction for GLM,  such as given by the values in Table 3 of Ref. 
\cite{Gotsman:2015aga}. In this case the values in parenthesis represent the model for which new parameters of their 
model were provided (Model IIn) - the ones describing the total, elastic and diffractive cross sections (low and high mass) 
at LHC. As emphasized by them that model gives higher values for the survival probability.
The spread in values given by the Telaviv group  depends on the impact parameter distribution of the hard amplitude, 
with a Gaussian  $e^{-b^2/4B}$ behaviour , which leads to the correct Froissart limit, and on different sets of parameters 
(called ``new" and ``old"), and also on the inclusion of kinematics  corrections. This work  develops in the context of  a  CGC/saturation approach for soft interactions at high
energy and is   detailed in  Ref.\cite{Gotsman:2015aga}.

On the other hand, it is rather clear what our present single-channel model can predict. 
The integrand in Eq.~(\ref{eq:survival}) depends on two quantities:
 \begin{enumerate}
\item $P_{no-inelastic}$ which is fixed by the fit to  the total cross-section, i.e. in single channel by the 
function $\chi_I(b,s)$ which describes $\sigma_{total}(s)$; in our interpretation, 
{ missing part or all of diffraction},  
$P_{no-inelastic }\equiv P_{only-diffractive-events}$
\item
the impact parameter distribution of partons which 
 can come from either soft collisions, with $A_{FF}(b)$,   or hard, mini-jet collisions, 
with $A_{BN}(b,s)$. In our  single  channel eikonal, we have seen in \cite{Grau:2009qx} that  
$A_{BN}(b,s)\sim e^{-(b\bar{\lambda})^{2p}}$. With the singularity parameter $1/2<p<1$ 
[our phenomenology indicates $p\simeq (0.6\div 0.7)$], one can see that the cut off in b-space is midway 
between a gaussian and an exponential,  leading to an asymptotic behavior of the total cross-section  
$\sigma_{total} \lesssim [\ln s]^{1/p}$, satisfying the Froissart bound. Please notice that an improvement  of  the Froissart limit in the context of the AdS/CFT correspondence has recently been proposed in  \cite{Diez:2015vha}.
\end{enumerate}

From this discussion, it is not completely clear which model  gives the best representation for the survival probabilities. 
While we are convinced that previous estimates of survival probabilities through mini-jet or QCD inspired models 
should be calculated using our proposal Eq.~(\ref{eq:survsum}) rather than Eq.~(\ref{eq:survBlock}) or (\ref{eq:survourold}),  
at the same time we are aware of the limitations of the single-channel model.  We expect that full exclusion of the 
hadronic background may further reduce the survival probability. 
\section{Final  comments and conclusions }
 The survival probability concept  implies the need to be able to select processes which are unaccompanied 
 by the usual hadronic activity. This may be useful for a selected process such as  Higgs production,  
 high $p_t$ jets, or any hard process which one wants to isolate from the background. The quantity to look 
 for is therefore a no-collision probability which is  characterized by the presence of rapidity gaps, around the central region.
Such quantity  is easily calculated   in the
 eikonal formulation.
 However,
  { the  single channel formulation of the inelastic cross-section given in Eq.~(\ref{eq:siginel})} fails 
 to reproduce the totality of the inelastic cross-section, as the energy increases towards LHC regime. 
 At energies lower than those attained at LHC, data for  the full inelastic cross-section are obtained by 
 subtraction from two well measured quantities, the total and the elastic, i.e. $\sigma_{inel}= \sigma_{total}-\sigma_{el}$. 
 This quantity includes events with different topologies, distinguished in various groups, such as soft, hard diffraction, 
 and central diffraction. The contribution from processes in the very forward direction is not uniquely 
 measurable by the different experiments, and  data are provided in terms of the covered phase space or through 
 model extrapolations. 

Here we propose that  the least ambiguous  way to use the concept of survival probability
is through  
selecting  events which do not have hadronic activity outside the diffractive region. In early release of elastic 
cross-section data at LHC, a common phase space limitation $\xi=M_X^2/s \ge 5 \times 10^{-6}$ was shown 
to be well described by a single channel model such as ours. Therefore one can now turn this fact around and 
define this region as the one for which the 
{ present  single channel BN model can provide an estimate for the 
survival rapidity gaps in the central region.} Namely, if ${\bar n}(b,s)\equiv 2\chi_I(b,s)$ is chosen so as to describe the total cross-section, then
\begin{equation}
P_{no-inel}(b,s)=e^{-{\bar n}(b,s)} \label{eq:Pno-inel}
\end{equation}
gives the probability distribution for   no independently distributed collisions at impact parameter  value $b$, 
and given c.m. energy. The survival probability at any given impact value $b$ is then dependent on the density of 
partons in the overlapping area in b-space. For central collisions, the hadronic matter is denser (confinement 
dilutes  gluonic matter in the peripheral regions) and {\it vice versa} for the peripheral collisions. By integrating the 
probability of no collisions with the hadronic matter distribution in the hadron, we have calculated the   
survival probability [for the case in which  Single
Diffraction is  not excluded] and found an 
estimate of $\sim1\%$ at LHC8 and LHC13. 

{The result we have presented obtains through an all order resummation procedure applied to soft gluon emission in mini-jet collisions.}

{Finally, we notice that results similar to the ones we are proposing for the BN model, can be expected in the model of Block {\it et al.} \cite{Block:2001ru} when this model is applied using our prescription.}
\begin{acknowledgments}
One of us, G.P., acknowledges hospitality at the MIT Center for Theoretical Physics, and is grateful to Earle Lomon for stimulating discussions and drawing our attention to the comparison with experimental data. Enlightening discussions with  Rohini Godbole are gratefully acknowledged.  We also acknowledge a  stimulating conversation with Valery Khoze about estimates of survival probabilities in different rapidity regions and for specific experimental cuts, such as for the CMS and ATLAS diffractive dijet data presented in the text. Y.S. thanks the Department of Physics and Geology at the University of Perugia for the continued
hospitality and acknowledges interesting discussions with O. Panella, L.
Fano' and S. Pacetti. A.G. acknowledges partial support by the
 Ministerio de Econom\'\i a y Competitividad 
(Kingdom of Spain),
 under grant number FPA2016-78220-C3-3-P, 
and by Consejer\'\i a de Econom\'\i a, Innovaci\'on, Ciencia y Empleo, Junta de Andaluc\'\i a, 
 (Kingdom of Spain),
(Grants FQM 101 and FQM 6552). D.A.F. is grateful to the INFN  Frascati National
Laboratories (LNF) for hospitality. 
\end{acknowledgments}
  \appendix
  \section{Derivation of semi-classical resummation formula }
  \label{app:etp}
   
  In this Appendix, we present a derivation of Eq.~(\ref{eq:dp2k}) following \cite{Etim:1967aa}. In the scattering of high energy charged particles, a considerable portion of the energy is radiated away in the form of of either hard or  soft  radiation,  photons in QED, gluons in QCD. In this appendix, we shall outline the method of soft quanta resummation we use in QCD.
  
  When a charged particle is bent in its path by the electromagnetic field, the cloud of soft photons accompanying the motion of the charged particle is not affected by the external field and continues its path, tangent to the trajectory at the point  when the charged particle  entered the bending field. This is true in QED where the external field has no effect on the soft photons surrounding the traveling electron, but it is more difficult to see it in QCD. However, this is true also in this case, because of the infrared catastrophe, as can be seen through a reading of  the Block and Nordsieck theorem  \cite{Bloch:1937pw} which demonstrates that only the emission of an infinite number of soft photons has a finite probability. This argument can be  applied as well to soft gluon emission, being   based on  ignoring the recoil of the emitting particles and summing of  all the Poisson distributed number of soft quanta. Thus in the emission, when the gluon energy goes to zero, the number of emitted gluons is infinite and the emitted energy accompanying this infinite number of gluons is color neutral and the soft gluon cohort does not interact with the external field. 
  
  In the language of quantum field theory, we can describe the process of soft gluon emission as a process in which straightforward perturbation theory does not apply in the sense that only the emitted energy-momentum is a finite quantity, not the single soft gluons. In order to resum this infinite number of soft gluons (or soft quanta emitted by a charged source) we can start from the theory of emission from a classical source as given by  Bloch and Nordsieck, in which the distribution of soft photons is shown to be given by a Poisson distribution,  i.e.
  \begin{equation}
  P(\{ n_{\bf k}\})=\Pi_{\bf k}\frac{
  [{\bar n}_{\bf k}]
  ^{n_{\bf k}}
  }{
  n_{\bf k}!
  } 
  e^{
  -{\bar n}_{\bf k}
  }
  \end{equation}
where $P(\{ n_{\bf k}\})$ can now be taken to  corresponds to the probability that the emitted massless quanta are emitted with $n_{{\bf k}_1} $ gluons with momentum ${\bf k}_1$,  $n_{{\bf k}_2} $ gluons with momentum ${\bf k}_2$, etc. The next step is to consider the overall energy-momentum loss $K$ accompanying the scattering and impose energy-momentum conservation to the sum of all the possible Poisson distributions, i.e.
\begin{equation}
d^4P(K)=\sum_{n_{\bf k'}} P(\{ n_{\bf k'}\})\delta^4(\sum_{k'} k' {n_{\bf k'}} -K)d^4K
\end{equation}  
The sum over all the distributions runs  again along the lines of a classical derivation, using the  four-dimension integral representation of the $\delta$ function, which allows to exchange the order of product of distributions and their summation. One thus reaches the expression
\begin{equation}
d^4P(K)=\frac{d^4 K}{(2\pi)^4} 
\int d^4 x\  e^{-h(x)+iK\cdot x} \label{eq:dP4k}
\end{equation}
with
\begin{equation}
h(x)=\sum_k(1-e^{-ik\cdot x}){\bar n}_{\bf k}\label{eq:hx}
\end{equation}
Going from the discrete to the continuum limit and integrating Eqs.~(\ref{eq:dP4k}) and(\ref{eq:hx}) on the unobserved variables of energy $K_0$ and longitudinal momentum $K_3$,  one obtains Eq.~(\ref{eq:dp2k}). The derivation can be applied to gluons or photons, provided the resulting integrand in Eq.~(\ref {eq:hpp})
be an integrable function. In QED  this quantity is not just  integrable but is also finite. The QCD limit is discussed in the text, with the proposal, put forward in \cite{Nakamura:1984na}, that  the integrand in Eq.~(\ref {eq:hpp}) be singular but integrable. This leads to  the condition that the infrared limit of the soft gluon coupling to the emitting source be no more singular than $(k_\perp^2)^{-p}$ with $p<1$ and  to the adoption of this limit in  the phenomenological approach, which  we have called   the BN model.

\section{The full inelastic cross-section from the empirical model}\label{app:empirical}
For the case when background emission in a wider phase space has to be excluded,   
a simple way to estimate the full inelastic cross-section can be obtained from the empirical model 
of \cite{Fagundes:2013aja}. We present here the results of  this model, although we shall not use  
it to estimate the survival probabilities, in absence of a clear indication of how to  calculate the impact parameter 
distribution of partons to associate to this model. 
 We  consider 
an empirical model based on the improved parametrization of the elastic amplitude  following   
the Phillips and Barger \cite{Phillips:1974vt}  proposal. As TOTEM data  \cite{Antchev:2011zz} for 
the differential elastic cross-section appeared, we discussed the validity of  this model in \cite{Grau:2012wy} and, 
 in \cite{Fagundes:2013aja}, we revised it,  proposing two different modifications, labelled $mBP1$ and $mBP2$, 
 aimed to ameliorate the description of the amplitude at $t=0$, and obtain a better fit of the total cross-section. 
 
 Our improved expression
  is  based on a best fit  to all  $pp$ differential cross-section data from ISR energies up 
   to $\sqrt{s}=7$ TeV, using a parametrization  of the elastic amplitude, which, in
     the  $mBP2$ version of the  empirical model,  was proposed to be
   \begin{equation}
    {\cal A}(s,t)=i[F^2_P(t/t_0)\sqrt{A(s)}e^{B(s)t/2}+e^{i\phi(s)}\sqrt{C(s)}e^{D(s)t/2}] \label{eq:mpb2}
   \end{equation}
where $   F^2_P(t)$ is the square of the proton form factor, i.e. $F_{P}(t/t_0)=1/[(1+|t|/t_0)]^2$ with $t_0$ a parameter with weak energy dependence, approaching $0.7 \ GeV^2$ at high energies.  
{The introduction of this factor in the first term at the right hand side of  Eq. ~(\ref{eq:mpb2})  modifies the  Phillips and Barger proposal to  give a better agreement with total cross-section data.} 

This  model  has 6 real parameters:  two amplitudes, $A(s)$ and $C(s)$,  two slopes,  $B(s)$ and  $D(s)$, one phase $\phi$ and one scale $t_0$. The model was able to give an excellent description of available data up to $\sqrt{s}=7$ TeV, and can be used to extrapolate to higher energies.  Using the full range of ISR and LHC7 data, we can  make predictions for the two amplitudes and the two slopes at higher energies by means of asymptotic theorems. As for the phase and the scale,
    { while the phase was kept constant,  for the energy dependence of $t_{0}(s)$ we use the interpolation/extrapolation fit result: $t_{0}=0.66+15.4/\log^{2}(s/1\ \text{GeV}^{2})$, which gives a good description of the $t_0$ parameter  in \cite{Fagundes:2013aja}.}
    The   values we propose to be used for the parameters in the LHC energy range are given in Table \ref{tab:empirical}.

 Using  the amplitude of Eq. (\ref{eq:mpb2}) and the asymptotic projections for the two amplitudes $A(s),\  C(s)$ and the two slopes $B(s)$ and $D(s)$,   we   calculate  the total cross-section at much higher than present energies,  and compare it with data. And then,  always from the above amplitude, one can also calculate and predict values for the elastic total cross-section, and, by default, for the inelastic, $\sigma_{inel}^{emp}=\sigma_{tot}^{emp}-\sigma_{el}^{emp}$. These expectations are shown as dotted lines in Fig.~\ref{fig:alltogether}.

 We see that  both the elastic  and the total cross-section are   well described by the empirical model parametrization at all energies - in fact the 8 TeV TOTEM values for the total cross-section is correctly predicted to be 103 mb vs. the TOTEM value at 102.9 mb - while   the inelastic cross-section appears  slightly higher than the TOTEM data and clearly higher than CMS.
 
 \begin{table}[htb]
\centering
\caption{Energy evolution of the mBP2 empirical model parameters used in Fig.~\ref{fig:alltogether}. We have assumed a nearly constant phase, $\phi\simeq 2.9$ rad, throughout in our calculations.\label{tab:empirical} }
\vspace{+0.5cm}
{\scriptsize
\begin{tabular}{|c||c|c|c|c|c||}
\hline \hline 
$\sqrt{s}$ (\text{TeV}) & A (\text{mbGeV}$^{2}$) & B (\text{GeV}$^{-2}$) & C (\text{mbGeV}$^{2}$) & D (\text{GeV}$^{-2}$) &$ t_{0}$ (\text{GeV}$^{2}$) \\ 
\hline 
0.5 & 197 & 4.83 & 0.217 & 3.19 & 0.760 \\ 
\hline 
2.0 & 344 & 6.50 & 0.693 & 4.00 & 0.727 \\ 
\hline 
8.0 & 597 & 8.78 & 1.30 & 4.80 & 0.708 \\  
\hline 
13 & 719 & 9.71 & 1.51 & 5.08 & 0.703 \\ 
\hline 
20 & 846 & 10.6 & 1.69 & 5.33 & 0.699 \\ 
\hline 
50 & 1180 & 12.6 & 2.06 & 5.87 & 0.693 \\ 
\hline \hline 
\end{tabular} }
\end{table}

\bibliography{diff_data_2}
\end{document}